\documentclass[a4paper]{ESASPCS13Style}
\usepackage{epsfig}

\begin{document}

\title{Submillimetre continuum emission from Class\,0 sources: Theory, Observations, and
Modelling}

\author{M. Rengel\inst{1}, K. Hodapp\inst{2}, D. Froebrich\inst{3}, S. Wolf\inst{4}, and J. Eisl\"offel\inst{1}}
  \institute{Th\"uringer Landessternwarte Tautenburg, Sternwarte 5, 07778 Tautenburg,
  Germany
\and Institute for Astronomy, 640 N. A'ohoku Place, Hilo, HI 96720, USA
\and Dublin Institute for Advanced Studies, 5 Merrion Square, Dublin 2, Ireland
\and Max Planck Institute for Astronomy, K\"onigstuhl 17, 69117 Heidelberg, Germany }

\maketitle

\begin{abstract}

We report on a study of the thermal dust emission of the
circumstellar envelopes of a sample of Class\,0 sources. The
physical structure (geometry, radial intensity profile, spatial
temperature and spectral energy distribution) and properties (mass,
size, bolometric luminosity ($L_{\rm bol}$) and temperature ($T_{\rm
bol}$), and age) of Class\,0 sources are derived here in an
evolutionary context. This is done by combining SCUBA imaging at
450 and 850~$\mu$m of the thermal dust emission of envelopes
of Class\,0 sources in the Perseus and Orion molecular cloud
complexes with a model of the envelope, with the implementation of
techniques like the blackbody fitting and radiative transfer
calculations of dusty envelopes, and with the Smith evolutionary
model for protostars. The modelling results obtained here confirm
the validity of a simple spherical symmetric model envelope, and
the assumptions about density and dust distributions following the
standard envelope model. The spherically model reproduces reasonably
well the observed SEDs and the radial profiles of the sources. The
implications of the derived properties for protostellar evolution
are illustrated by analysis of the $L_{\rm bol}$, the $T_{\rm bol}$,
and the power-law index $p$ of the density distribution for a
sample of Class\,0 sources.

\keywords{Stars: formation, circumstellar matter, low-mass --
radiative transfer -- dust --submillimeter  \ }
\end{abstract}

\section{Introduction}

The earliest evolutionary phase of star formation, in which a deeply
embedded protostar is known to exist, is the so-called Class\,0
stage (Andr\'e et al.~1993). Since protostars in this phase 
(Class\,0 sources) are highly obscured by extended dusty envelopes,
they can be observed mainly in the far-infrared to millimetre
wavelength range. Specifically, a Class\,0 source consists of a
central protostellar object surrounded by an infalling dusty
envelope and a flattened accretion disk. Many (presumably all)
Class\,0 sources are associated with strong molecular flows, usually
more energetic than those from Class\,1 sources (Davis \&
Eisl\"offel~1995).

Since Class\,0 sources will determine most of the early evolutionary
history of the protostars as they evolve toward main-sequence stars, it
is crucial to determine the physical conditions within the envelope
of Class\,0 objects. However, because of detection difficulty,
constraining physical properties of Class\,0 sources is a
challenging task. Nevertheless, continuum submillimetre observations
of these sources let us to detect dust emission of the massive
circumstellar envelopes, and provide a powerful tool for
constraining the distribution of matter (Adams~1991).

Independent methods of detection of Class\,0 sources, and techniques
and useful ways to help to the interpretation of properties of
protostars have been developed and used during the last years (e.g.
comparation of submillimetre with near infrared surveys of a star
forming region (Stanke et al.~2000), interpretation from the dust emission (Shirley et al.~2000), blackbody
fitting (Chini et al.~2001), radiative transfer codes (Wolf et al.~1999), evolutionary models for
protostars (Smith~2002)). In this work, we combine them and develop an approach
in order to constrain the structure (geometry, radial intensity
profile, spatial temperature, and spectral energy distributions
(SEDs)), and physical properties (mass, size, bolometric luminosity
($L_{\rm bol}$) and temperature ($T_{\rm bol}$), and age) of Class\,0
sources. Specifically, we combine the information from the dust
emission with a physical model of the extended envelope, the
implementation of techniques like the blackbody fitting and the 1D
continuum self-consistent MC3D radiative transfer code (by Wolf et
al.~1999), and the Smith evolutionary model for protostars (by
Smith~2002).

To describe and study stellar evolution, the Hertzsprung -Russell
Diagram (HRD) has become the main way. However, to study
protostellar evolution, the $T_{\rm bol}$--$L_{\rm bol}$
diagram (BLTD, bolometric luminosity-temperature diagram) was
proposed as direct analogue to the HRD (Myers \& Ladd~1993; Myers et
al.~1998). There, Class\,0 sources are characterised by the lowest
$T_{\rm bol}$ values of all protostars. Although the BLTD provides a
useful way to study the evolution of protostars for all evolutionary
phases of relevance for star formation (Myers et al.~1998),
nevertheless, the Smith scheme predicts the evolution of $T_{\rm
bol}$ and $L_{\rm bol}$ as an envelope of initial mass $M_{0}$ and
temperature $T$ dissipates on a timescale $t$ and forms a star.

\section{Theory}
\label{theory}

How can we describe the protostellar emission from the envelope? In
order to facilitate the interpretation of submillimetre
observations, we adopt the standard envelope model (Adams~1991) to
keep the problem as simple as possible. The circular symmetry of the
observed emission (see Rengel~2004), and the lack of significant
internal structure justifies the simplicity of a spherical model
case. If the emission is optically thin, the observed intensity for
a spherically symmetric protostellar envelope at an impact parameter
$b$ is given by Eq.\,1 assuming a single power-law index of the
opacity $\beta$.

\begin{equation}\label{i}
I_{\nu}(b)=2\kappa_{\nu}\int^{r_{0}}_{b}
B_{\nu}\left[T_{d}(r)\right]\rho(r)\frac{r}{\sqrt{r^{2}-b^{2}}}dr,
\end{equation}

\noindent $r_{0}$ is the outer radius,
$\rho$ the density, $T_{d}$ the dust temperature, $\kappa_{\nu}$ the
opacity of the dust grains, and $B_{\nu}[T_{d}(r)]$ the Planck
function at dust temperature $T_d$. 

The current theory suggests that the density and temperature distributions
are power-law indices ($p$ and $q$, respectively, see e.g. Adams~1991). 
If the emission is in the
Rayleigh-Jeans limit and if $r_{0}$$\gg$ $b$, Eq.\,1 can be
approximated to $I_{\nu}(b)/I_{\nu}(0)=(b/b_{0})^{-m}$ (where $m$ is
the power-law index of the observed intensity profile, and $b_{0}$
is the normalization factor to the peak emission).

%

\section{Observations}
\label{observations}

To investigate the physical structure, processes, and properties of
Class\,0 sources, six star forming regions were observed in the
Perseus and Orion molecular cloud complexes at 450 and
850~$\mu$m with the Submillimetre Common-User Bolometer Array
(SCUBA) camera at the James Clerk Maxwell Telescope (JCMT) on Mauna
Kea, Hawaii. The target regions were \object{L1448}, \object{L1455},
\object{NGC\,1333}, \object{HH211}, \object{L1634}, and
\object{L1641\,N}. A distance of 300 pc for L1448 is assumed
according to Bachiller \& Cernicharo~\cite{bc86b}, 350~pc for L1455,
NGC\,1333 and HH211 (Chandler \& Richer~2000, and Herbig \&
Jones~1983), 460~pc for L1634 (Bohigas et al.~1993), and 390~pc for
L1641 (Anthony-Twarog~1982).

\begin{figure}[ht!]
\centering
 \includegraphics[angle=-90,width=8.2cm, bb=60 70 500 530]{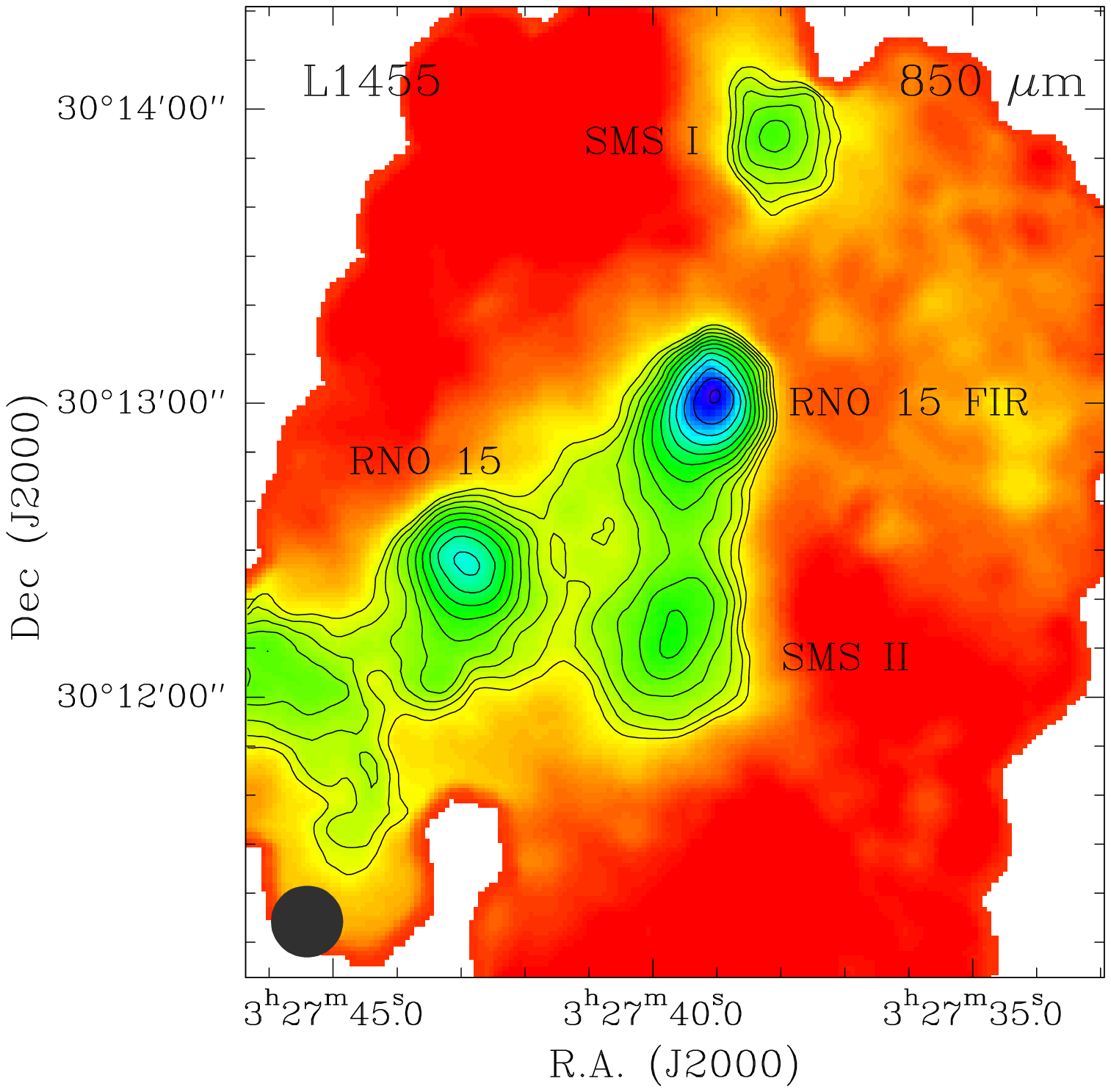}\\
\includegraphics[angle=-90,width=8.2cm, bb=60 70 500 530]{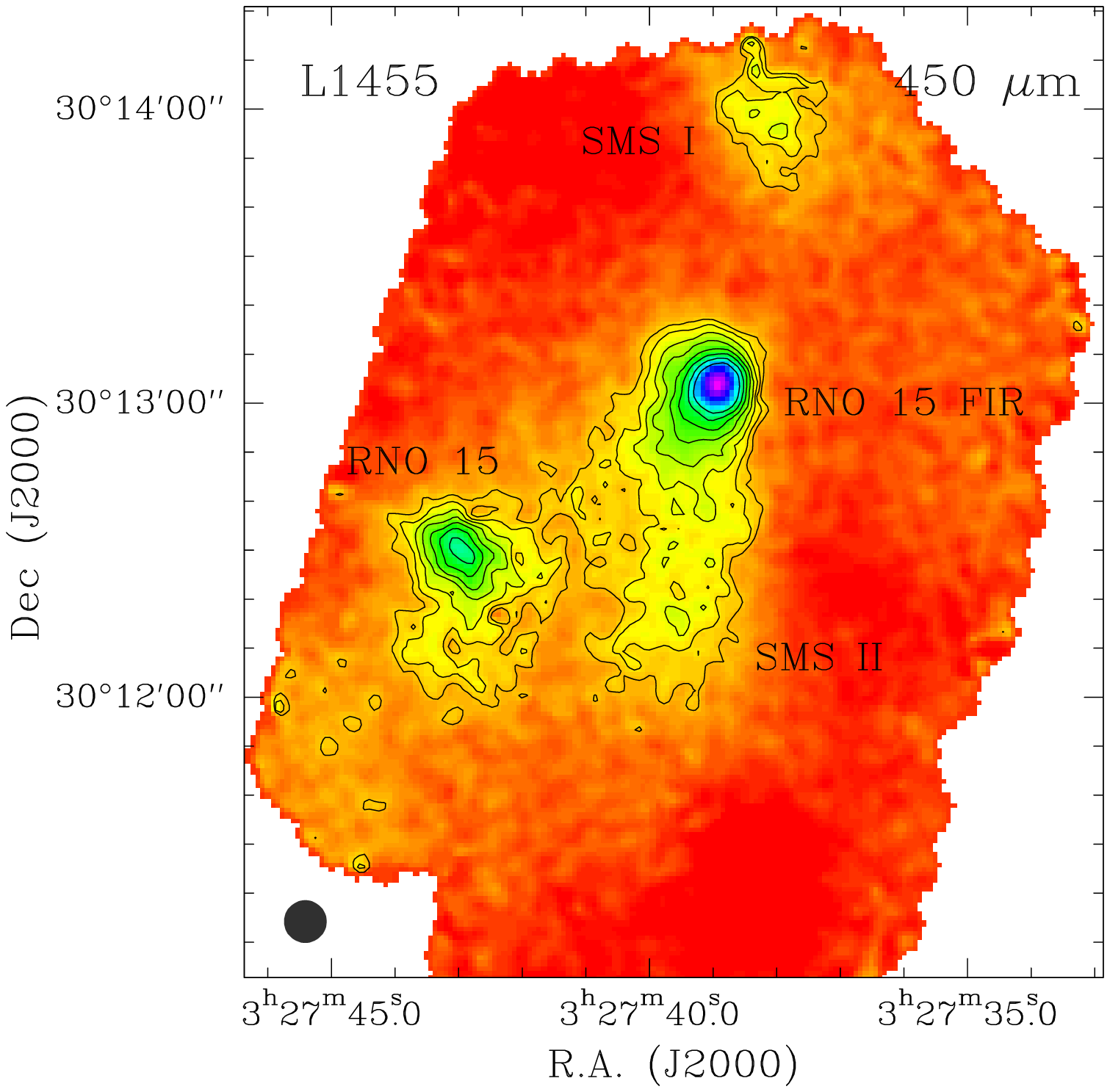}
\caption{Maps of the region L1455, overlaid with contours. The top
map is taken at 850~$\mu$m and the bottom map at 450~$\mu$m. In L\,1455 the
protostars \object{RNO\,15\,FIR} and \object{RNO\,15}, and the
pre-stellar cores SMS\,I and II are observed. Contour levels are in
log scale with step size 1.15 from 0.0005~Jy/beam at 850~$\mu$m, and
step size 1.25 from 0.0035~Jy/beam at 450~$\mu$m. The filled circles
at the bottom left indicate the main beam size.\label{fig1}}
\end{figure}

As an example, the resulting combined SCUBA maps at 450 and
850~$\mu$m of L1455 overlaid with isophotal contour plots are
shown in Fig.\,\ref{fig1}.

From these continuum SCUBA maps, 36 submillimetre sources ($>$
3$\sigma$) are detected (26 in Perseus and 10 in Orion). Some of
them are extended, and many contain multiple condensations, as well
as extended diffuse features. Twelve objects are reported for the first
time (six in Perseus and six in Orion), and fifteen sources are
analyzed here. Additionally, the first detailed submillimetre
observations of the two regions NGC\,1333 South and L1641\,N have been obtained and
are discussed in Rengel~(2004), unveiling objects previously
unknown. The survey also reveals a large number of small groups
(Rengel et al.~2003).

\subsection{Derivation of
parameters from the emission}

Physical parameters like the dust submillimetre spectral index
$\alpha_{450/850}$ (this value is related directly to the dust
emissivity exponent $\beta$ by $\beta$=$\alpha$ - 2), gas and dust
masses, and sizes of the envelopes are derived. The dust
submillimetre spectral index $\alpha_{\lambda_{1}/\lambda_{2}}$ is
defined by

\begin{equation}
\alpha_{\lambda_{1}/\lambda_{2}}=\frac{lo
g(S_{\lambda_{1}}/S_{\lambda_{2}})}{log(\lambda_{2}/\lambda_{1})}.
\end{equation}

The mean value of $\alpha_{450/850}$ for these objects is 2.8 $\pm$
0.4, implying that the regions investigated here are quite cold
($\sim$10\,K). Such low dust temperatures have for example also been
found for isolated globules in the Perseus region (see e.g. Myers \&
Benson~1983), suggesting an exponent $\beta \sim\,1$ for the dust
grain opacity, $\kappa_{\nu}$. The failure of the Rayleigh-Jeans
approximation could increase the value of $\beta$.

An aspect ratio was defined for each of the sources by dividing
the semi-major and minor axis lengths as measured typically at the
97\% contour. We find that the observed aspect ratios of the protostellar 
emission maps lie in the range 1.0-2.0 with an average value of 1.3 $\pm$ 0.1,
suggesting that the envelopes of
Class\,0 sources can be described approximately as spherically
symmetric. Nevertheless, a departure of this symmetry, i.e., from
an aspect ratio of 1, in some cases is found. This mismatch could
be produced by several factors like magnetic fields of energy
densities sufficiently large to influence the object structure and
cause flattening along field lines, the bipolar outflow, rapidly
rotating structures, or high levels of angular momentum (Rengel~2004).

\begin{figure}[t!]
  \begin{center}
    \epsfig{file=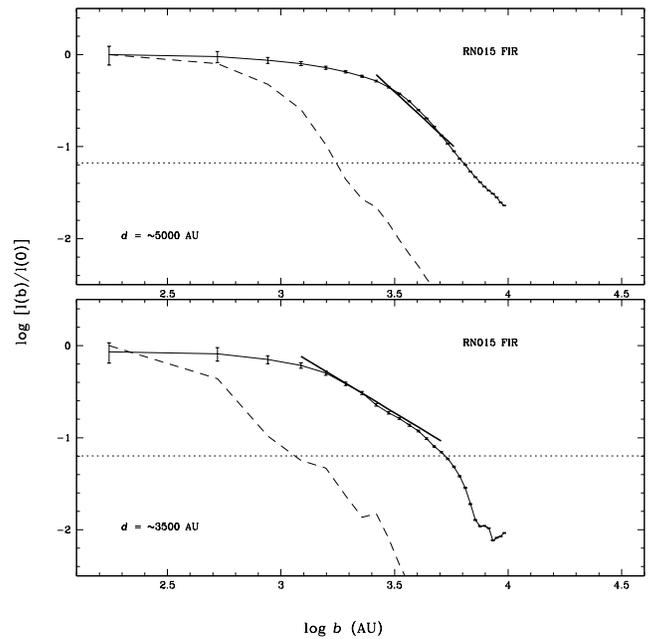, width=0.5\textwidth}
  \end{center}
\caption{Radial profiles of \object{RNO\,15\,FIR} at 850 $\mu$m (top
panel) and 450 $\mu$m (bottom panel). The normalized intensity is
plotted as a function of impact parameter, $b$(AU). The bold solid
lines represent the fitted slope $m$. The horizontal dashed lines
represent the 1\,$\sigma$ noise level at the edge of the map. The
beam profiles are shown as dashed lines. 
Derived diameters $d$ of the source (in AU) are given in the lower
parts.\label{fig2}}
\end{figure}

The average value of the gas and dust mass of the sample, based on
the thermal emission from the dust according to Hildebrand~(1983),
is 2.5 $\pm$ 0.6 M$_{\odot}$ (typical aperture 45$''$). In order to
derive physical properties of the sources from the observed
azimuthally averaged radial intensity profiles of the thermal
emission from the dust, like the power-law index of the temperature
distribution as a function of the radius $T$($r$) $\propto r^{-q}$,
$q$, $p$, and the envelope sizes, the standard envelope model
(Adams~1991) was adopted. Assuming that the observed sources can be
fitted by a power-law intensity distribution
$I_{\nu}(b)/I_{0}=(b/b_{0})^{-m}$ as Class\,0/1 sources, the
average value of $<$$m$$>$ is $1.74\pm 0.02$ at 850~$\mu$m and $1.50
\pm 0.02$ at 450~$\mu$m.
Following the analytic estimate for the intensity (Adams~1991),
these values of $<$$m$$>$ correspond to the expected ones for very
young objects: theoretical models as well as numerical simulations
of the collapse of an isothermal sphere (e.g., Shu~1977;
Larson~1969; Penston~1969) also predict power-law indices of the
density distribution, $\rho$($r$) $\propto r^{-p}$, with $p \leq$
2.0. Considering $m$=$p$ + $q$ -1, the first two order terms of the
equivalent power-law index $\mu$ of the observed intensity
(Adams~1991), the values $q$= $0.42\pm0.04$ as well as $p$=
$2.1\pm0.1$ and 2.3$\pm$0.1 (at 450 and 850~$\mu$m,
respectively) are found. In Fig.\,\ref{fig2}, the normalized radial profiles
[$I_{\nu}$($b$)/$I_{\nu}$(0)] (being $I_{\nu}$(0) the observed peak flux) of RNO\,15\,FIR as an example are
plotted versus $b$.

To estimate the source size, an outer diameter $d$ corresponding to
the measured Half Power Beam Width (HPBW) size is adopted. The
observed sources are surrounded by extended envelopes, having
typical sizes of 1500-6000~AU (at 450~$\mu$m) and 4000-9000~AU (at
850~$\mu$m).

\section{Modelling}
\label{modelling}

\subsection{Blackbody fitting}
\label{bb}

Accurate SEDs are computed for nine sources combining existing
multi-wavelength surveys with the submillimetre data obtained in
Sect.\,3 (Rengel~2004), and fitting a modified blackbody curve to
the fluxes (Froebrich et al.~2003). This method computes a grid of
different blackbody curves of the form (\ref{formalsed}) in which
the temperature, the optical depth at 100 $\mu$m and $\beta$ are
varied ($15$\,K $\leq$\,$T$\,$\leq$ $80$\,K; $0.0 \leq \beta \leq
3.0$; $0.09 \leq \tau_{100} \leq 11.4$). Specifically, curves of the form  (\ref{formalsed}) are 
convolved with the filter curves of the used filter bands
(Eq.\,\ref{fluxfilter}), and compared to the data-points. The model that best fits the
observational data is determined. We derived the
bolometric temperature and luminosity, size of the envelope, and
sub-millimetre slope for each source.

In local thermodynamic equilibrium, the solution of the equation of
transfer for an isothermal object that subtends an effective solid
angle $\Sigma \Omega$ is given by the emergent flux density $S$:

\begin{equation}
S[Jy]= \Sigma \Omega \cdot \left(1-e^{-\tau}\right) \cdot B(\lambda,
T_{d}). \label{formalsed}
\end{equation}

\noindent $B(\lambda,T$) denotes the Planck function
for a dust temperature $T_{d}$, and $\tau$ the dust optical depth.
The optical depth can be expressed by

\begin{equation}
\tau=\tau_{100} \cdot \left(\frac{\lambda}{100\,{\rm
\mu\,m}}\right)^{-\beta}. \label{tau100}
\end{equation}

\noindent $\lambda$ is in $\mu$m, the optical depth at 100 $\mu$m ($
\tau_{100}$) is a free parameter, and $\beta$ is the submillimetre
slope of the SED ($\beta$=0 corresponds to a blackbody).

But we do not observe the emergent flux density $S_{\lambda}$
directly, we observe $S_{f}$, the emergent flux detected across a
filter $f$, which has a filter transmission curve $T_{f}(\lambda)$.
$S_{f}$ is calculated by

\begin{equation}
S_f[Jy] = \frac{\Sigma \Omega \, \int\limits_{\lambda = 0}^{\infty}
\left(1-e^{-\tau}\right) \, B(\lambda, T) \, T_f(\lambda) \,
d\lambda}{\int\limits_{\lambda = 0}^{\infty} T_f(\lambda) \,
d\lambda} \label{fluxfilter}
\end{equation}

\noindent for each filter $f$ separately using the filter
transmission curves $T_{f}(\lambda)$. 
For details about how the calculations of the maximum likelihood $L$ were performed, see Froebrich et al.~(2003).

The SEDs revealed that the sample consists of cool objects ($T_{\rm
bol}$ ranges of $\sim$27-50~K) which have $L_{\rm bol}$ of
$\sim$4-85~L${_\odot}$. Thus, by means of the sub-millimetre to
bolometric luminosity ratio, \object{L1448\,NW}, \object{L1448\,C},
\object{RNO\,15\,FIR}, \object{NGC 1333\,IRAS\,1}, \object{NGC\,1333\,IRAS\,2}, \object{HH211-MM},
\object{L\,1634}, \object{L 16 41\,N}, and \object{L1641 SMS\,III} are
Class\,0 sources. For detailed computations of the associated errors
to each determined parameters, see Rengel~(2004). In Fig.\,\ref{fig4} (bottom,
short dashed lines), best obtained fit to the data-points using our
approach and overplotted the combined data of \object{RNO 15\,FIR}
are shown as an example.

Because of the IRAS 12 and 25~$\mu$m points are usually far
above the fit to the SED (e.g. Chini et al.~2001; Barsony et
al.~1998), and of the unclear quantification of a higher bolometric temperature value caused by filter leaks,
outflow/dust interactions or envelope emission if these IRAS points are included on SED computations ($\sim$4~K
Froebrich et al.~2003), 
these points are excluded on the target list and on the blackbody fits performed here.
In Fig.\,\ref{fig4}, the blackbody curve and the data-points show significant
deviations at the Infrared Astronomy Satellite (IRAS) 12 and
25~$\mu$m points for the Class\,0 sample. If these data-points
had correct flux measurements, the sources would exhibit an excess
of mid-infrared emission which could be caused by ongoing
outflow/dust interactions. Because it is likely that these IRAS
points are incorrect in some cases, further mid-infrared
observations of these objects are necessary in order to investigate
the nature of this possible excess.

\subsection{MC3D radiative transfer code}
\label{mc3d}

Results from the standard theory provide the first direct insights
into observable estimations (e.g. $p$ and $q$, Sect.\,3.1). Nevertheless, the
temperature profile will diverge from a single power-law $q$ as the
envelope becomes optically thick at the primary wavelengths of
energy transport (inner portion of the envelope) (e.g. Shirley et
al.~2000). Here it becomes necessary to calculate the temperature
distribution self-consistently by implementation of a radiative
transfer code.

\begin{figure}[b!]
  \begin{center}
    \epsfig{file=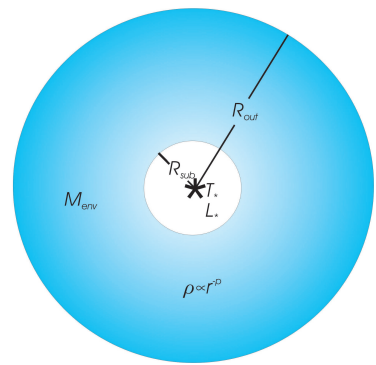, width=0.27\textwidth}
  \end{center}
\caption{Basic model assumed in the MC3D code.\label{fig3}}
\end{figure}

Detailed physical interpretation of the internal structure of nine
objects is realized by intensive computer radiative transfer
modelling. In particular, dust temperature distributions, SEDs, and
intensity maps are derived self-consistently by the MC3D code (Wolf
et al.~1999). The basic model assumed is: a central heating source
that is embedded within a spherical symmetric dust envelope of mass
$M_{\rm env}$. The central source is characterized by a stellar
luminosity ($L_{*}$), and an effective temperature ($T_{*}$). The
dust envelope is specified by an outer radius ($R_{\rm out}$), a
density power-law index $p$, an inner radius of the envelope $R_{\rm
in}$, which is set by the dust destruction radius $R_{\rm sub}$
(also called sublimation radius) adopted to a temperature of
1500\,K, the dust destruction temperature of graphite (e.g. Adams \&
Shu~1985, which occurs at a temperature of 1500~K; Myers et
al.~1998). The central source luminosity $L_{*}$ is the sole source
of luminosity for the envelope, so $L_{*}$=$L_{\rm bol}$.
Fig.\,\ref{fig3} shows a scheme of the adopted model. The validity
of this model, and physical parameters of the sources (e.g. envelope
masses, density distributions, sizes, and sublimation radius) are
derived by finding the consistency between observed and modelled
SEDs, and radial profiles (Sect\,4.3; Rengel et al.~2004).

The dominant opacity source in the radiative transfer models is dust
grains. The characteristics of grains in Class\,0 sources are
unknown. However, graphite and silicate grains have been proposed as
likely consistent of Galactic dust. The grain distribution adopted
here consists of small interstellar grain size 0.005-0.25 $\mu$m,
with mixing of astronomical silicate and graphite (relative
abundances: 62.5 \% \& 37.5 \% (respectively)), and the Draine \&
Lee~(1984) optical constants to describe grain properties. The
adopted model assumes that the dust properties are not a function of
radius in the envelope. The gas-to-dust mass ratio is adopted to be
100:1, 
whereby we use a dust grain density of 3.6~g~cm$^{-3}$. The wavelength grid
(60 wavelengths) are chosen to cover the relevant range.

\begin{figure}[t!]
\centering
\includegraphics[width=0.4\textwidth, bb=15 150 600 750]{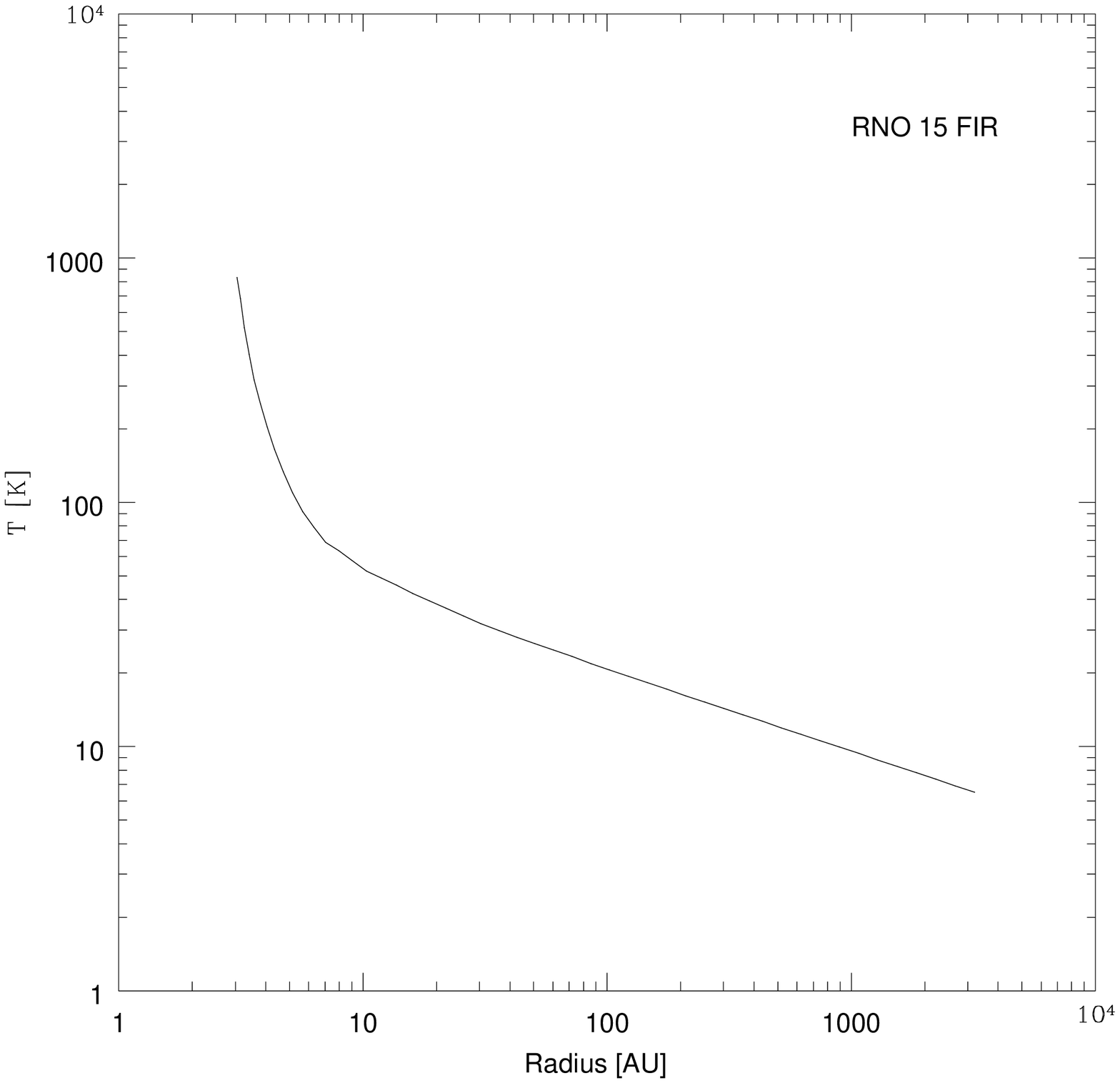}
\includegraphics[width=0.4\textwidth, bb=15 150 600 750]{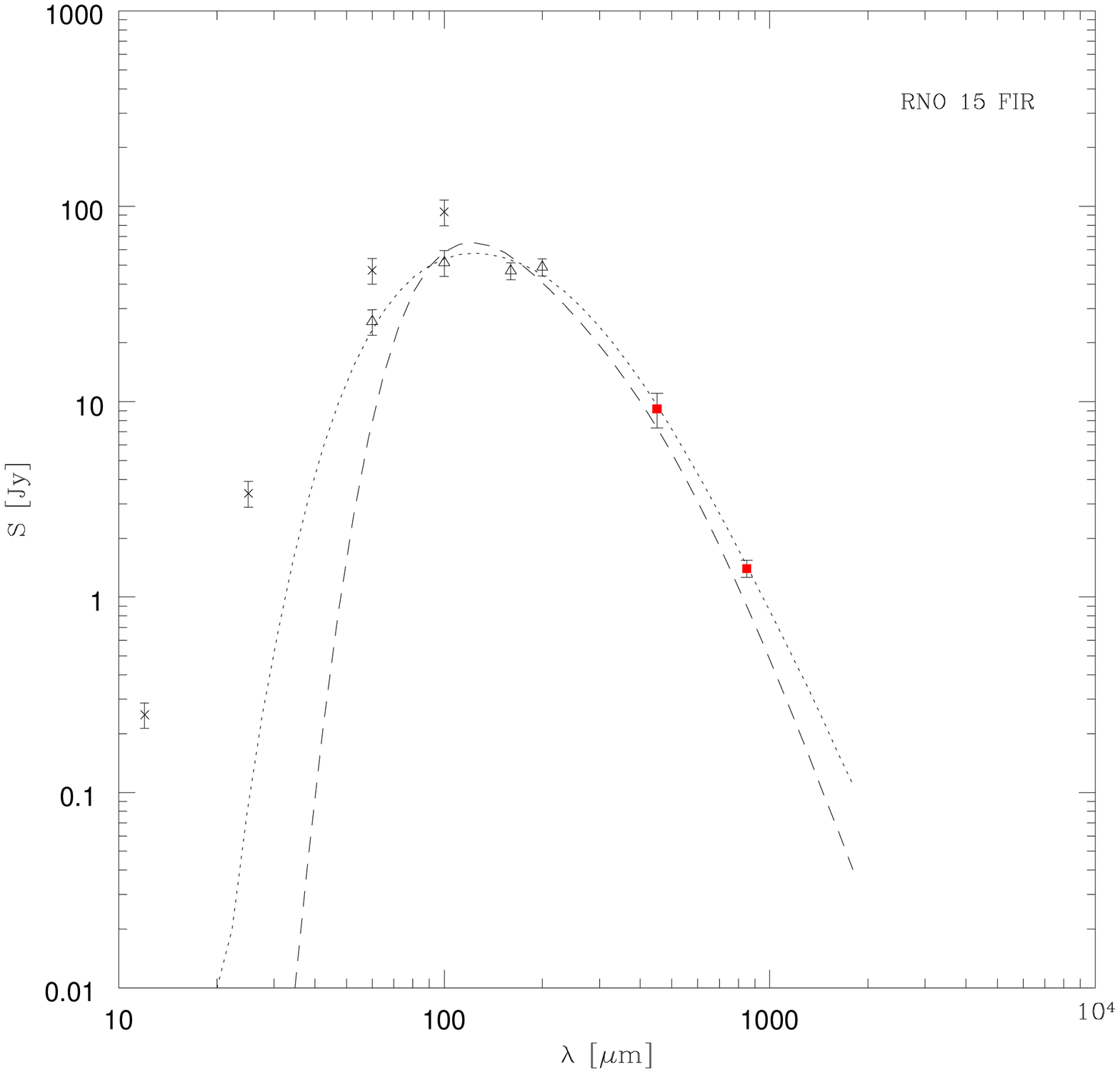}
\caption{\label{fig4} {Temperature profiles for RNO\,15\,FIR
performed using the MC3D radiative transfer code
(Wolf et al.~2003) (top). Best simulated and observed SEDs (bottom,
short dashed and dotted lines, respectively). The filled squares
represent the measured 450 and 850~$\mu$m fluxes from the
SCUBA observations presented in Sect.\,3. The IRAS 12 and
25~$\mu$m points are excluded from the fit of the SED. Details
regarding the summarized target list at several wavelengths from the
literature are given in Rengel~(2004). For the model results, see
Table\,1.}}
\end{figure}

In order to derive the physical structure of the envelope, three
steps are carried out: the first step constrains the spatial
temperature structure of the circumstellar envelope, the second step
determines the simulated SED, and the third step, derives the
emission, which provides the intensity structure. These last
observables lead to test the model. At this point it is necessary to
keep in mind that the values of the inferred parameters are valid
only within the framework of the model assumptions considered here.
In Fig.\,\ref{fig4}, the temperature profile, and best simulated and observed
SEDs for RNO\,15\,FIR are shown as an example.

The central source should radiate from a specific radius $r$, where
the layer changes between the optically thick and the thin regime.
This radius $r$ is called ``photospheric radius" or ``effective
photosphere", which depends on the wavelength and the molecule
species. Crudely, the observed emission arises from a
``characteristic" layer at $\tau_{\lambda}$=1 (Myers et al.~1998) or
2/3 (Kenyon et al.~1993; Hartmann~1998). The temperature
distribution changes its character inside the photospheric radius,
where the envelope is optically thick; trapping of radiation
causes the temperature gradient to become steeper than in the
optically-thin case. To calculate the photospheric radius, $\tau$=1
and the spherically symmetric optically thick envelope of Myers et
al.~(1998), with an inner radius $R_{\rm in}$, a density $\rho_{\rm
in}$ and an outer radius $R_{\rm out}$, are adopted here for
simplicity. Following this model,
the photospheric radius $r$ is given by
\begin{equation}
\label{pr} r_{\tau=1}\approx R_{\rm in} \tau_{12},
\end{equation}
\noindent $\tau_{12}$ being the optical depth from $R_{\rm in}$ to
$R_{\rm out}$.

Following the Myers et al.~(1998) model, a sublimation radius of
3-5~AU at 450~$\mu$m, $\tau$=1 and $p$=2 ($\sim$10~AU for $p$=1.5)
is estimated, which is roughly the same as the radius of the
photosphere.

As result of the modelling, it is found that the physical structure
of a Class\,0 envelope is characterized by the gas and dust
temperature distribution $T$($r$) $\propto r^{-q}$, and the density
distribution $\rho$($r$) $\propto$ $r^{-p}$, with $q$=0.4 and $p$ in
the range of 1.5-2. In the inner part of the envelope (10~AU),
$T$($r$) departs from the single-power law index of 0.4, resulting
in a steeper inner temperature profile. This phenomenon can be
produced probably by thermal convection in the inner envelope and
causing zones of dust destruction by evaporation when the envelope
becomes optically thick (e.g. Lin \& Papaloizou~1980). $T$ accounts
for the effects of this heating.

\subsection{Comparison of models with observations}
\label{mo}

We estimate the parameters for the Class\,0 sources by
comparing the observed SEDs obtained by blackbody fitting
(Sect.\,\ref{bb}) and the simulated SEDs from the envelope fitting
procedure (Sect.\,\ref{mc3d}). Following the same technique as Kenyon et al.~1993, 
the best model for a given observational SED is
determined by minimizing the normalized sum of the weighted squared
residuals,

\begin{equation}
S^2=\sum w_i(F_{o,i}-F_{m,i})^{2}/F_{o,i}
\end{equation}

\noindent where $F_{o,i}$ is the observed flux at wavelength $i$,
$F_{m,i}$ the scaled flux of the model, and $w_{i}$ a
weighting factor\footnote{a weighting factor is empirically given to
the consistency between the modeled and observed SED, with
dependence on the wavelength. Reasonable weights are given from the
uncertainties in each region of the SED.}. Given the uncertainties
in estimating observational errors and the limitations of the
models, a more complicated fitting treatment is unjustified. Because
of the uncertainties in opacity, the difficulties in matching
observational beam size with model results, and the possible
contribution of disks to the submillimetre region (c.f. Beckwith et
al.~1990; Butnet et al.~1991; Keene \& Masson~1990), we used weights
$w_i$=1 for wavelengths 60 $\mu$m $\leq$ $\lambda$ $\leq$ 300~
$\mu$m, and chose $w_i$=0.5 for 300~$\mu$m $<$ $\lambda$ $<$ 1000
$\mu$m. Data-points taken at wavelengths shorter than 40 $\mu$m are excluded in the comparation 
calculations because they are expected to be optically thick  
and therefore very sensitive to asymmetric geometries (e.g. outflow
cavities and flattened envelopes), which are currently unable to be
modelled. Because of the spectral range considered here does not extend much over long wavelengths,
data-points taken at wavelengths longer than 1000 $\mu$m are rejected.

\begin{figure}[Hb!]
\centering
\includegraphics[angle=-90,width=0.398\textwidth, bb=15 0 600 750]{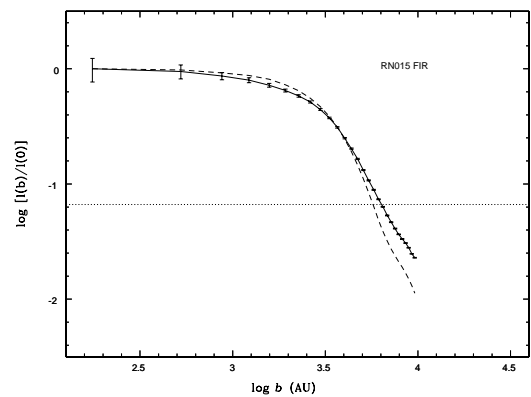}\hspace{0cm}
\caption{\label{fig5} {Simulated and observed radial profiles
as a function of the impact parameter $b$(AU) of RNO\,15\,FIR at 850
$\mu$m (short dashed and solid lines, respectively). The dotted
lines represent the 1\,$\sigma$ noise level at the edge of the
map.}}
\end{figure}

As it is claimed previously, a comparison between best simulated and
observed SEDs, and between simulated and observed radial profiles
led us to check the reliability
of the model. In order to create the simulated radial profile, the
envelope emission at 850~$\mu$m is calculated as an image from the
MC3D code for each source, and convolved by the SCUBA beam. Fig.\,\ref{fig5}
shows the observed and simulated radial profile of
\object{RNO\,15\,FIR} as an example.

We conclude from our comparison that the assumption of spherical
symmetry in the envelope is a reasonable approximation to study the
envelope of Class\,0 sources. Nevertheless, deviations from this
geometry between observed and modeled SEDs at short wavelengths, and
between observed and simulated radial profiles of relatively
spherical Class\,0 sources become extremely prominent (e.g. for
RNO\,15\,FIR and NGC\,1333\,IRAS\,1) (see Rengel~2004). In
addition, small differences between observed and simulated radial
profiles are observed for HH\,211-MM and RNO 15\,FIR (aspect ratios
of 1.2 and 1.3 at 850~$\mu$m, respectively). This discrepancy could
be due to the exclusion of a non-spherical geometry, of cavities
produced by molecular outflows, or of a flattened disk in the model;
it could also be due to specific assumptions considered here,
e.g.\,a single point-like nature of the sources (e.g. 
it is uncertain that HH211-MM and RNO\,15\,FIR are point-like sources
(Eisl\"offel et al.~2003; Davis et al.~1997, respectively)), and a
constant dust opacity with radius in the model.

\subsection{Bolometric luminosity-temperature diagram}
\label{bltd}

\begin{figure}[ht!]
  \begin{center}
    \epsfig{file=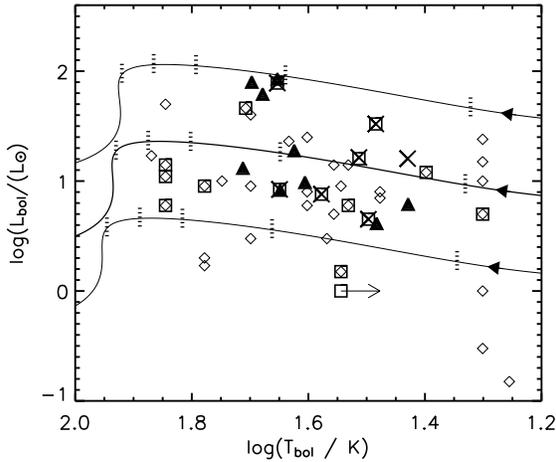, width=0.43\textwidth,}
  \end{center}
\caption{The BLTD for the objects analyzed here (filled triangles),
the Class\,0 data from Andr\'e et al.~(2000) (open
squares), from Froebrich et al~(2003) (thick Xs), and 
from Stanke~2000 (open diamonds).
Protostars evolve from right to the left. Three tracks for the final
masses of 0.2, 1 and 5~M$_\odot$ are displayed. The vertical dotted
lines on the tracks mark the model ages of 20, 30, 40, and 50
thousand years (tracks derived by combining the Unification Scheme
(Smith 2000; 2002) with the framework for protostellar envelopes
presented by Myers et al.~(1998)).} \label{fig6}
\end{figure}

The observed properties of Class\,0 sources can be understood more
thoroughly if it is assumed that the objects evolve in time. Using
the Smith protostellar evolutionary scheme (Smith~2002), and
constructing the $L_{\rm bol}$--$T_{\rm bol}$ diagram for the
sample, the age, the infall rate, and the envelope mass as a
function of time are estimated for the detected Class\,0 sources
(Rengel~2004). In particular, comparing the positions of the objects
with an evolutionary model age and mass of the envelope of each
source are derived (Fig.\,\ref{fig6}, filled triangles).

The tracks in Fig.\,\ref{fig6} are derived by combining the Unification
Scheme, as reviewed by Smith (2000, 2002), with the framework for
protostellar envelopes presented by Myers et al. (1998).

For the sample, the resulting model-dependent ages of Class\,0
sources are in a range from 1 to 3 $\times$ $10^{4}$ yr. The
evolution of these sources is traced by $T_{\rm bol}$, which
typically increases systematically between 25-50~K. The resulting bolometric
luminosities span a range between 5-85~L$_{\odot}$.

Following the Smith scheme (with an accretion rate not constant in
time, Smith~2002), the accretion rate $\dot M_{\rm acc}$ at the
earliest Class\,0 stage increases exponentially for a short interval
of time (with an accretion peak in 1.5$\times$10$^4$ yr), and then
declines later on (Rengel~2004).

\begin{figure}[ht]
  \begin{center}
    \epsfig{file=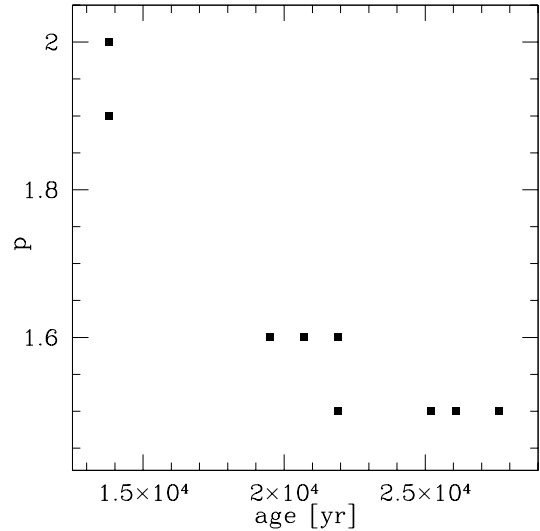, width=0.41\textwidth}
  \end{center}
\caption{The power-law index $p$ of the sample 
as function of age listed in Table\,1. Detailed
computations show that the errors of $p$ are less or equal to 0.1.}
\label{fig7}
\end{figure}

Is there a correlation of the power-law index $p$ with time? We plot
the estimated values of $p$ as a function of $t$. Fig.\,\ref{fig7} suggests a
density structure of $\rho$($r$) $ \propto r^{-2}$ at younger ages,
evolving to $\rho$($r$) $ \propto r^{-3/2}$ at later times. In order
to investigate this with a larger spread, a Class\,1 sample is
mandatory.

\section{Results}

Table\,1 lists the characteristics of the best-fitting model
parameters for the sample of sources analyzed in detail here.

The main results of this study are: 1) From the continuum SCUBA
maps, 36 submillimetre sources ($>$ 3$\sigma$) are detected. Some of
them are extended, and many contain multiple condensations, as well
as extended diffuse features. Twelve of these objects are reported
here for the first time, 15 sources are analyzed, and nine Class\,0
sources are studied in detail. 2) An average value of the aspect
ratio of the sample of 1.3 $\pm$ 0.1 is found, suggesting that the
envelopes of Class\,0 sources can be described approximately as
spherically symmetric. Nevertheless, a departure of this symmetry in
some cases is found. This mismatch could be produced by several
factors like magnetic fields of energy densities sufficiently large
to influence the object structure, the bipolar outflow, rapidly
rotating structures, or high levels of angular momentum. 3) The
sample consists of cool objects ($T_{\rm bol}$ ranges of
$\sim$27-50~K) which have $L_{\rm bol}$ of $\sim$4-85~L${_\odot}$.
4) The physical structure of a Class\,0 envelope is characterized by
the gas and dust temperature distribution $T$($r$) $\propto r^{-q}$,
and the density distribution $\rho$($r$) $\propto$ $r^{-p}$, with
$q$=0.4 and $p$ in the range of 1.5-2 ($p$=2 at younger ages and
$p$=1.5 at later times). In the inner part of the envelope (10~AU),
$T$($r$) departs from the single-power law index of 0.4, a
phenomenon caused probably by thermal convection in the inner
envelope. 5) From the thermal emission of the dust and following an
emissivity model (Hildebrand~1983), the average value of the gas and
dust mass of a sample based on 15 submillimetre selected objects is
2.5 $\pm$ 0.6~M$_{\odot}$. Examination of the radial profiles of the
sample shows that the objects are surrounded by extended envelopes.
Typically, the sizes are $\sim$1500-6000~AU (at 450~$\mu$m) and
$\sim$4000-9000~AU (at 850~$\mu$m). 6) A sublimation radius of
3-5~AU at 450~$\mu$m, $\tau$=1 and $p$=2 ($\sim$10~AU for $p$=1.5)
is estimated, which is roughly the same as the radius of the
photosphere (because the gas that is accreted close to the central
object is optically thick, the central source should radiate from
this specific radius, where the layer changes between the optically
thick and the thin regime). 7) The blackbody curve and the
data-points show significant deviations at the Infrared Astronomy
Satellite (IRAS) 12~$\mu$m and 25~$\mu$m points for the Class\,0
sample. If these data-points had correct flux measurements, the
sources would exhibit an excess of mid-infrared emission which could
be caused by ongoing outflow/dust interactions. Because it is likely
that these IRAS points are incorrect in some cases, further
mid-infrared observations of these objects are necessary to
investigate the nature of this possible excess. 8) The assumption of
spherical symmetry in the envelope model is reasonable.
Nevertheless, deviations from this geometry determined by comparations between 
observed and
modeled SEDs at short wavelengths, and between observed and
simulated radial profiles of relatively spherical Class\,0 sources
become extremely prominent. This discrepancy could be due to the
exclusion of a non-spherical geometry, of cavities produced by
molecular outflows, or of a flattened disk in the model; it could
also be due to specific assumptions considered here, e.g.\,a single
point-like nature of the sources, and a constant dust opacity with
radius in the model. 9) For the sample of Class\,0 sources, the
model-dependent ages are in a range of 1-3$\times$10$^{4}$~yr.

\section{Conclusions}

Detailed physical interpretation of the internal structure of nine
observed Class\,0 sources was carried out by intensive computer
modelling. A simple spherically symmetric model envelope, and
assumptions about density and dust distributions following the
standard envelope model reproduced reasonably well the observed SEDs
and the radial profiles of the sources. $T$($r$) $\propto r^{-0.4}$
is a good approximation for the sample. The radial temperature
distribution as function of distance, however, departed
significantly from the optically-thin assumption, an observational
derivation of a single-power law $q$ of 0.4 for radii $<$ 10~AU.
Modelling results indicated a density profile well described by a
power-law between $p$=1.5 and 2, which is expected by  

\begin{table*}[tH!]
  \caption{Best envelope fit results for the embedded sources. The temperature of the
central object T$_{*}$ is set to 3500~K. $\chi^{2}$ quantifies the
agreement between model and data. The age is estimated according to
the model of Smith (2000) and given in 10$^3$\,yrs.}
  \label{tab:table}
  \begin{center}
    \leavevmode
    \footnotesize
    \begin{tabular}[h]{lcccccccc}
      \hline \\[-5pt]
      Object & $T_{\rm bol}$ [K] & $L_*$ [L$_\odot$] &  $R_{\rm sub}$~[AU] & $p$ & $R_{\rm out}$~[AU] & age [$\times$
      10$^3$ yr]& $M_{\rm env}$\,[M$_{\odot}$] & $\chi^2$\\[+5pt]
      \hline \\[-5pt]
\object{L\,1448\,NW}              &27    &~6 &5        &2.0            &~4000  &13.8 &~2.8  &~6.4 \\
\object{L\,1448\,C}               &40    &11 &5        &1.6            &~3000  &19.5 &~1.7  &~2.6 \\
\object{RNO\,15\,FIR}             &45    &~8 &3        &1.6            &~3500  &20.7 &~1.0  &~1.8 \\
\object{NGC\,1333\,IRAS\,1}       &52    &13 &3        &1.5            &~4500  &21.9 &~1.5  &~8.6 \\
\object{NGC\,1333\,IRAS\,2}       &48    &68 &3        &1.5            &~7000  &26.1 &~3.0  &~0.1 \\
\object{HH\,211-MM}               &30    &~5 &3        &1.9            &~6000  &13.8 &~4.1  &~4.5 \\
\object{L\,1634}                  &42    &19 &3        &1.6            &~6000  &21.9 &~2.8  &~1.1 \\
\object{L\,1641\,N}               &45    &85 &3        &1.5            &~8500  &25.2 &~7.0  &~1.0 \\
L\,1641\,SMS\,III        &50    &80 &3        &1.5            &10000  &27.6 &~6.3  &~0.6 \\
     \hline \\
      \end{tabular}
  \end{center}
\end{table*}

\noindent all of the collapse models and numerical studies.
A density structure of
$\rho$($r$) $\propto r^{-2}$ at younger ages, evolving to
$\rho$($r$) $ \propto r^{-3/2}$ at later times was found.

SED and radial profile fits constrained physical parameters of the
sources such as envelope masses, density distributions, sizes and
sublimation radii. Nevertheless, mid-infrared observations, and the
inclusion of outflow and disk, and other geometries in the model are
desirable to investigate in order to decrease the differences
between the model and the data, or prove that data and model are
different under several configurations.

\begin{acknowledgements}

MR thanks financial support from the Deutsche Forschungsgemeinschaft
(DFG) grant number Ei 409/6-2 and the Graduierternf\"orderung of the
Friedrich-Schiller-Universit\"at.

\end{acknowledgements}

%
%

\end{document}